\documentclass[aps,pre,twocolumn,preprintnumbers,superscriptaddress]{revtex4}
\usepackage{color}
\usepackage{multirow}
\usepackage[normalem]{ulem}
\usepackage{amsmath}
\usepackage{enumerate}
\usepackage{amsfonts}
\usepackage{epsfig}
\usepackage{graphicx}
\usepackage[caption=false]{subfig}
\newcommand{\be}{\begin{equation}}
\newcommand{\ee}{\end{equation}}
\newcommand{\bea}{\begin{eqnarray}}
\newcommand{\eea}{\end{eqnarray}}

\definecolor{hookersgreen}{rgb}{0.0, 0.44, 0.0}
\definecolor{patriarch}{rgb}{0.5, 0.0, 0.5}
\definecolor{awesome}{rgb}{1.0, 0.13, 0.32}
\definecolor{egyptianblue}{rgb}{0.06, 0.2, 0.65}
\definecolor{darkgreen}{rgb}{0,0.5,0}
\definecolor{darkblue}{rgb}{0,0,0.6}
\definecolor{purple}{rgb}{0.4,.2,0.7}
\usepackage[colorlinks=true,citecolor=darkgreen,linkcolor=purple,urlcolor=purple]{hyperref}

\def\le{\left}
\def\ri{\right}
\def\nr{N_{\rm r}}
\def\Q{Q}
\def\qzero{q_0}
\def\qone{q_1}

\begin{document}

\title{Glassy slowdown and replica-symmetry-breaking instantons}

\preprint{MIT-CTP-4552}

\author{Allan Adams}
\email{awa@mit.edu}
\affiliation{Center for Theoretical Physics, Massachusetts Institute of Technology,
Cambridge, Massachusetts 02139, USA}

\author{Tarek Anous}
\email{tanous@mit.edu}
\affiliation{Center for Theoretical Physics, Massachusetts Institute of Technology,
Cambridge, Massachusetts 02139, USA}

\author{Jaehoon Lee}
\email{jaehlee@mit.edu}
\affiliation{Center for Theoretical Physics, Massachusetts Institute of Technology,
Cambridge, Massachusetts 02139, USA}

\author{Sho Yaida}
\email{sho.yaida@duke.edu}
\affiliation{Center for Theoretical Physics, Massachusetts Institute of Technology,
Cambridge, Massachusetts 02139, USA}
\affiliation{Department of Chemistry, Duke University, Durham, North Carolina 27708, USA}

\begin{abstract}
Glass-forming liquids exhibit a dramatic dynamical slowdown as the temperature is lowered.
This can be attributed to relaxation proceeding via large structural rearrangements whose characteristic size increases as the system cools.
These cooperative rearrangements are well modeled by instantons in a replica effective field theory, with the size of the dominant instanton encoding the liquid's cavity point-to-set correlation length.
Varying the parameters of the effective theory corresponds to varying the statistics of the underlying free-energy landscape.
We demonstrate that, for a wide range of parameters, replica-symmetry-breaking instantons dominate.
The detailed structure of the dominant instanton provides a rich window into point-to-set correlations and glassy dynamics.
\end{abstract}
\maketitle
When glass-forming liquids are cooled over a modest range of temperatures, their dynamics slows down by many orders of magnitude.
This slowdown manifests itself in a rapid increase of shear viscosity and structural relaxation time.
Despite decades of work, the physics of the slowdown -- also observed in a wide variety of systems including granular and biological systems -- has yet to be properly described~\cite{BBreview,KTreview}.
We develop a method with which to fill this gap, and, in so doing, we find nonperturbative effects triggered by replica-symmetry-breaking (RSB) instantons, making predictions for correlations in glassy systems and opening up additional research directions for understanding glassy dynamics.

Glassy slowdown is believed to be controlled by growing cavity point-to-set (PTS) correlations~\cite{PTS_BB}, whose correlation length diverges together with the relaxation time: this has been rigorously established for a large class of graphical models~\cite{PTS_MS}, and there are numerical simulations that support this hypothesis for glass-forming liquids~\cite{PTS_BBCGV,PTS_HMR}.
To define cavity PTS correlations, consider a many-body system at equilibrium.
First, specify a cavity, say a spherical ball of size $R$, and pin everything outside of it -- thus fixing a \emph{set}.
Now randomize the particles inside the cavity and allow them to re-equilibrate under the influence of the force exerted by the external pinned particles.
The cavity PTS correlator measures the overlap between the new configuration and the original at a \emph{point} inside the cavity. 
When the cavity is sufficiently small, the interior configuration will be strongly constrained by the fixed configuration outside the cavity, so the cavity PTS correlation should be large.
In contrast, when the cavity is sufficiently large, the two configurations become statistically independent deep in their interiors.
This crossover from high correlation to low defines the cavity PTS correlation length, $\xi_{\rm PTS}$.

The physics of cavity PTS correlations can be captured by an effective theory of a replica field $q_{ab}\le({\bf r}\ri)$, where $q_{ab}\le({\bf r}\ri)=q_{ba}\le({\bf r}\ri)$ and $q_{aa}\le({\bf r}\ri)=0$, with $a,b=1,2,...,\nr$~\cite{SY}.
The original equilibrium configuration singles out a replica index, say $a=1$, which acts as a fictitious disorder for re-equilibrated configurations inside the cavity.
The field components $q_{1 \tilde{a}}(\bf r)$ with $ \tilde{a}\ne1$ then characterize the position-dependent overlap between the original and the re-equilibrated configurations, while the others characterize the overlap between two independent re-equilibrated samples immersed in the fictitious disorder.
Pinning the external particles means that the overlap $q_{1 \tilde{a}}(\bf r)$ must be large outside the cavity.  Inside small cavities, the pinned boundary conditions keep the field in the high-overlap metastable state throughout.  Inside large cavities, in contrast, the boundary conditions cannot prevent the field from finding a low-overlap minimum near the core.
This crossover, taking place at $\xi_{\rm PTS}$, is precisely what an instanton captures, in the limit $\nr\rightarrow1$.
The dominance of large replica-field instantons indicates the need for large cooperative rearrangements in order for the system to continue sampling its phase space, resulting in sluggish dynamics.  

The effective-field-theoretic approach allows us to explore cavity PTS correlations in generic glassy systems by writing down a generic effective action incorporating all interactions in the $q_{ab}$ that are symmetric under permutations of the $\nr$ indices~\cite{SY}.
We start with the action~\cite{footnoteWolynes,footnoteparameter}
\bea\label{simple}
S[q_{ab}\le({\bf r}\ri)]&=&\frac{1}{2}\int d^3{\bf r} \sum_{a,b=1}^{\nr}\Bigg[\frac{1}{2}\le(\nabla q_{ab}\ri)^2+\frac{t}{2}q_{ab}^2-\frac{w}{3}q_{ab}^3\nonumber\\&&+\frac{y}{4}q_{ab}^4-\frac{u}{3}\le(q_{ab}^3+\sum_{c=1}^{\nr}q_{ab}q_{bc}q_{ca}\ri)\Bigg],
\eea
and we see how varying the parameter $u$, which couples different components of the replica field, changes the character of the dominant instanton.
While this action is not completely general, it is sufficient for demonstrating that RSB in PTS correlations is generic, in the following sense: turning on more general couplings such as $\sum_{a,b,c=1}^{\nr}q_{ab}q_{bc}$ may trigger higher-step RSB~\cite{truncatedSG}, but it does not change the fact that replica symmetry is generically broken.

\begin{figure}[t]
\centering
\includegraphics[width=8.6cm]{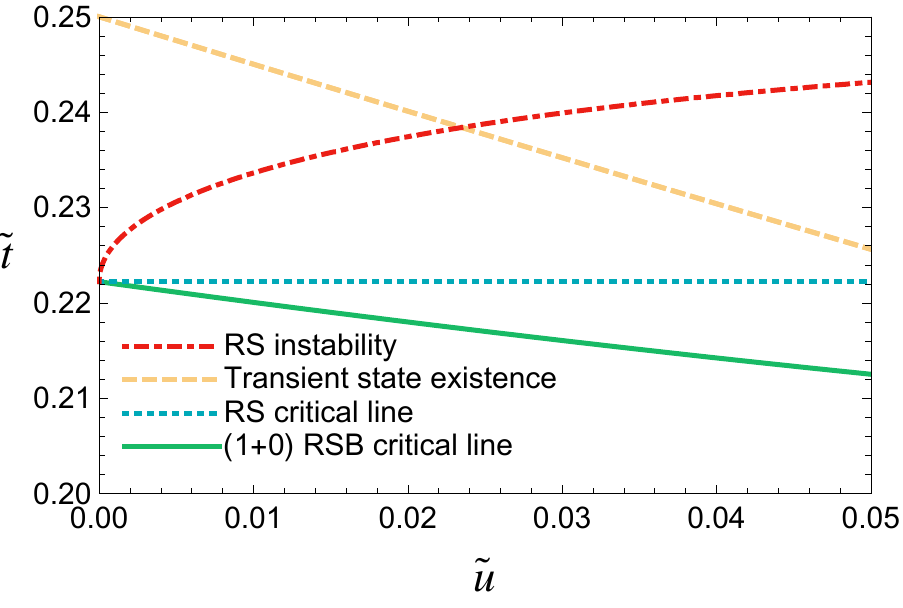}
\caption{
\textbf{Stability and critical lines.}
We plot various important lines in the space of dimensionless couplings $\tilde{t}\equiv t\, y/w^2$ and $\tilde{u}\equiv u/w$ in the regime explored.
Below the red dash-dotted curve the (1+0)-step RSB instanton dominates over the RS instanton.
The blue dotted line at $\tilde{t}=2/9$ indicates where the size of the RS instanton diverges.
The green solid line depicts the critical line where the size of the (1+0)-step RSB instanton diverges.
Below the yellow dashed line, the effective potential has a ``transient state''  local extremum, leading to the two-walled structure of the (1+0)-step RSB instanton.}
\label{stabilities}
\end{figure}

Since the effective action treats the replicas symmetrically, one might naively expect that the dominant instantons do not distinguish between replica indices.
Solutions of this type are called replica-symmetric (RS), and we begin by analyzing them.
Varying the action (\ref{simple}) gives the saddle-point equations for $q_{ab}\le({\bf r}\ri)$,
\be\label{general}
-\nabla^2q_{ab}+tq_{ab}-wq_{ab}^2+yq_{ab}^3  =  u\le[q_{ab}^2+\sum_{c=1}^{\nr}q_{ac}q_{cb}\ri]
\ee
for $a\ne b$. Inserting the RS ansatz $q_{ab}\le({\bf r}\ri)=(1-\delta_{ab})\Q\le({\bf r}\ri)$ into this saddle-point equation and taking $\nr\rightarrow1$ gives
\be\label{trivialEOM}
-\nabla^2\Q+t\Q-w\Q^2+y\Q^3 = 0\, .
\ee
This equation can be derived from a reduced action
\bea\label{trivial}
S_{\rm red}\le[\Q\le({\bf r}\ri)\ri]&\equiv&\lim_{\nr\rightarrow1}\frac{2 S[q_{ab}\le({\bf r}\ri)]}{\nr(\nr-1)}\Big|_{\rm RS}\nonumber\\
&=&\int d^3{\bf r}\ \le[\frac{1}{2}\le(\nabla \Q\ri)^2+V(Q)\ri]\, ,
\eea
with the potential term
\be
V(Q)=\frac{t}{2}\Q^2-\frac{w}{3}\Q^3+\frac{y}{4}\Q^4\, .
\ee
An RS instanton $\Q^\star({\bf r})$ is a solution to Eq.~(\ref{trivialEOM}), which asymptotes to the metastable minimum of the potential $V(Q)$, $Q=Q_{\rm meta}\equiv \frac{w+\sqrt{w^2-4yt}}{2y}$, at spatial infinity and approaches the trivial minimum, $Q=0$, towards the core. The dominant RS instanton is spherically symmetric~\cite{CGM}, and we can construct it numerically (see the discussion below on the RSB instanton).
Note that when $t=t^{\rm RS}_{\rm c}\equiv\frac{2w^2}{9y}$, the two minima have the same energy, implying that the size of the RS instanton diverges~\cite{ColemanAspects}.
Were the RS instanton the dominant solution in this parameter regime, its diverging size would indicate a diverging PTS correlation length. As we shall see, the RS instanton develops an instability {\em above} $t=t^{\rm RS}_{\rm c}$ (see Fig.~\ref{stabilities}).  Thus $t^{\rm RS}_{\rm c}$ overestimates the {PTS}-critical value of $t$.

To test the stability of the RS instanton, we consider infinitesimal perturbations around the RS solution, $q_{ab}\le({\bf r}\ri)=(1-\delta_{ab})\Q^\star\le({\bf r}\ri)+\delta q_{ab}\le({\bf r}\ri)$, and we compute the eigenvalues of the resulting Hessian.
The eigen-perturbations $\delta q_{ab}\le({\bf r}\ri)$ of the Hessiean can be sorted into three categories~\cite{AT,Nishimori}.
The first consists of perturbations that are symmetric among the replicas,
\be
\delta q_{ab}\le({\bf r}\ri)=(1-\delta_{ab})\delta\phi_{\rm I}\le({\bf r}\ri)\, .
\ee
For these modes, the remaining eigenvalue problem is, taking $\nr\rightarrow1$,
\be\label{first}
\le[-\nabla^2+t-2w\Q^\star+3y\Q^{\star 2}\ri]\delta\phi_{\rm I}^{(n)}=\lambda_{\rm I}^{(n)}\delta\phi_{\rm I}^{(n)}\, .
\ee
The second category consists of perturbations that single out one replica index.  For example, singling out the first index, we have
\bea
\delta q_{1\tilde{a}}\le({\bf r}\ri)&=&\le(\frac{2-\nr}{2}\ri)\delta \phi_{\rm II}\le({\bf r}\ri)\label{secondclass}\\
{\rm and}\ \ \ \delta q_{\tilde{a}\tilde{b}}\le({\bf r}\ri)&=&(1-\delta_{\tilde{a}\tilde{b}})\delta \phi_{\rm II}\le({\bf r}\ri)
\eea
for $ \tilde{a},\tilde{b}=2,3,...,\nr$.
The corresponding eigenvalue equation is, in the limit $\nr\rightarrow1$,
\be\label{second}
\le[-\nabla^2+t-\le(2w-u\ri)\Q^\star+3y\Q^{\star 2}\ri]\delta\phi_{\rm II}^{(n)}=\lambda_{\rm II}^{(n)}\delta\phi_{\rm II}^{(n)}\, .
\ee
Finally, the third category consists of perturbations that single out two distinct replica indices.
For example, singling out the first two indices, we have
\bea
\delta q_{12}\le({\bf r}\ri)&=&\le\{\frac{(2-\nr)(3-\nr)}{2}\ri\} \delta \phi_{\rm III}\le({\bf r}\ri)\, ,\\
\delta q_{1\check{a}}\le({\bf r}\ri)&=&\delta q_{2\check{a}}\le({\bf r}\ri)=\le(\frac{3-\nr}{2}\ri) \delta \phi_{\rm III}\le({\bf r}\ri)\, ,\\
{\rm and}\ \ \ \delta q_{\check{a}\check{b}}\le({\bf r}\ri)&=&(1-\delta_{\check{a}\check{b}})\delta \phi_{\rm III}\le({\bf r}\ri)
\eea
for $\check{a},\check{b}=3,4,...,\nr$.
The resulting eigenvalue equation is equivalent to Eq.~(\ref{first}).

For the first and third categories,
the eigenvalue equation is precisely the one that arises from the reduced action (\ref{trivial}).
In particular, the eigenvalue spectrum is independent of $u$.
These perturbations never represent an instability of the RS instanton.

For the second category, the story is more interesting, as can be seen by numerically evaluating the eigenvalue spectrum of Eq.~(\ref{second}) (see the Appendix).
As we vary the couplings, the lowest eigenvalue associated with the spherically symmetric mode sometimes crosses zero, signaling a real instability of the RS instanton.
In Fig.~\ref{stabilities}, the red dash-dotted curve indicates where this instability sets in.
Below this curve, the dominant instanton must break the replica symmetry.
But what is the dominant saddle?

The fact that the instability singles out one replica direction suggests a different ansatz, 
namely $q_{1\tilde{a}}\le({\bf r}\ri)=\Q\le({\bf r}\ri)$ and $q_{\tilde{a}\tilde{b}}\le({\bf r}\ri)=(1-\delta_{\tilde{a}\tilde{b}})\qzero\le({\bf r}\ri)$ for $\tilde{a},\tilde{b}=2,...,\nr$, where we single out the first replica index without loss of generality.  
Schematically,
\be\label{N0RSBpic}
q_{ab}\le({\bf r}\ri)=\left[ {\begin{array}{ccccc}
   0 & \Q & \Q & ... &\Q  \\
   \Q & 0 & \qzero & ... & \qzero  \\
   \Q & \qzero & 0 & ... & \qzero  \\
   ... & ... & ... & ... & ...\\
   \Q & \qzero & \qzero & ...& 0  \\
  \end{array} } \right]\le({\bf r}\ri)\, .
\ee
We will refer to this as the (1+0)-step RSB ansatz~\cite{footnotenaming}.

Plugging this ansatz into the saddle-point equations (\ref{general}) and taking $\nr\rightarrow 1$ gives
\be\label{N0RSB_EOM_1}
-\nabla^2\Q+t\Q-w\Q^2+y\Q^3 = -u\le[\,\Q\qzero-\Q^2\ri]
\ee
and
\be\label{N0RSB_EOM_2}
-\nabla^2\qzero+t\qzero-w\qzero^2+y\qzero^3 = -u\le[\,\qzero^2-\Q^2\ri]\, .
\ee
Since the mode that triggers RSB is spherically symmetric, we assume that the dominant (1+0)-step RSB instanton depends only on the radial coordinate $r=|{\bf r}|$.
Our problem can then be treated as a one-dimensional boundary value problem (BVP), with the boundary conditions that the fields asymptote to the metastable value $Q_{\rm meta}$ at spatial infinity $r=\infty$ and that their first derivatives vanish at the origin $r=0$ to ensure regularity of the solution.
We  numerically solve the resulting equations using the pseudospectral method, expanding the fields in a basis of Chebyshev polynomials, and solving the resulting nonlinear system via Newton iteration~\cite{SpectralMatlab} (see the Appendix).
To generate an initial guess that converges to the nontrivial (1+0)-step RSB instanton, we use a ``mountain pass" algorithm, which forces the fields to traverse barriers of the effective potential.
For computational efficiency, we employ a weakly adaptive mesh refinement algorithm to focus on regions where the gradients of the fields are large.

\begin{figure*}[t]
\centering
\subfloat[\ \ \ \ \ \ \ \ \ \ \ \ (a)]{
\includegraphics[width=8cm]{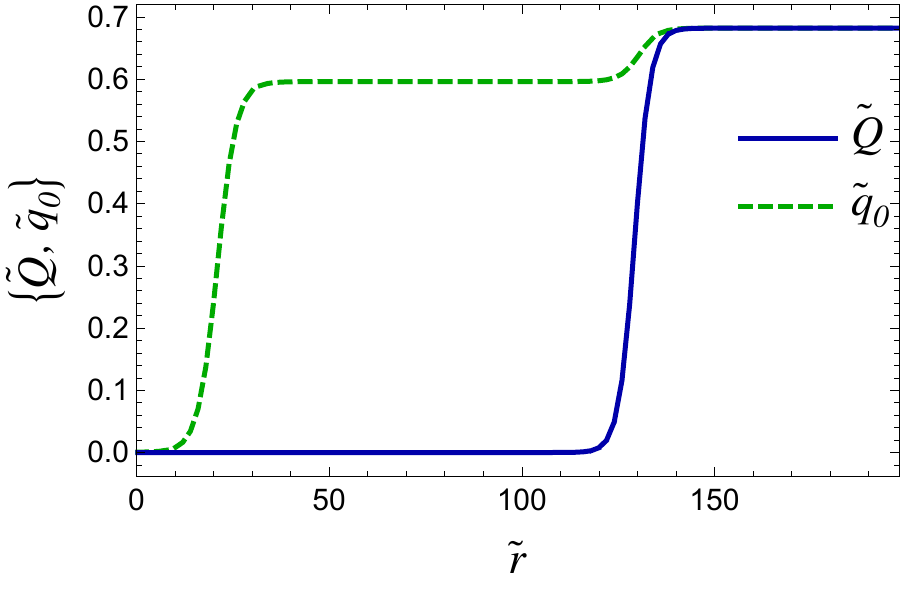}\label{instantonsb}}
\quad
\subfloat[\ \ \ \ \ \ \ \ \ \ \ (b)]{\includegraphics[width=9.3cm]{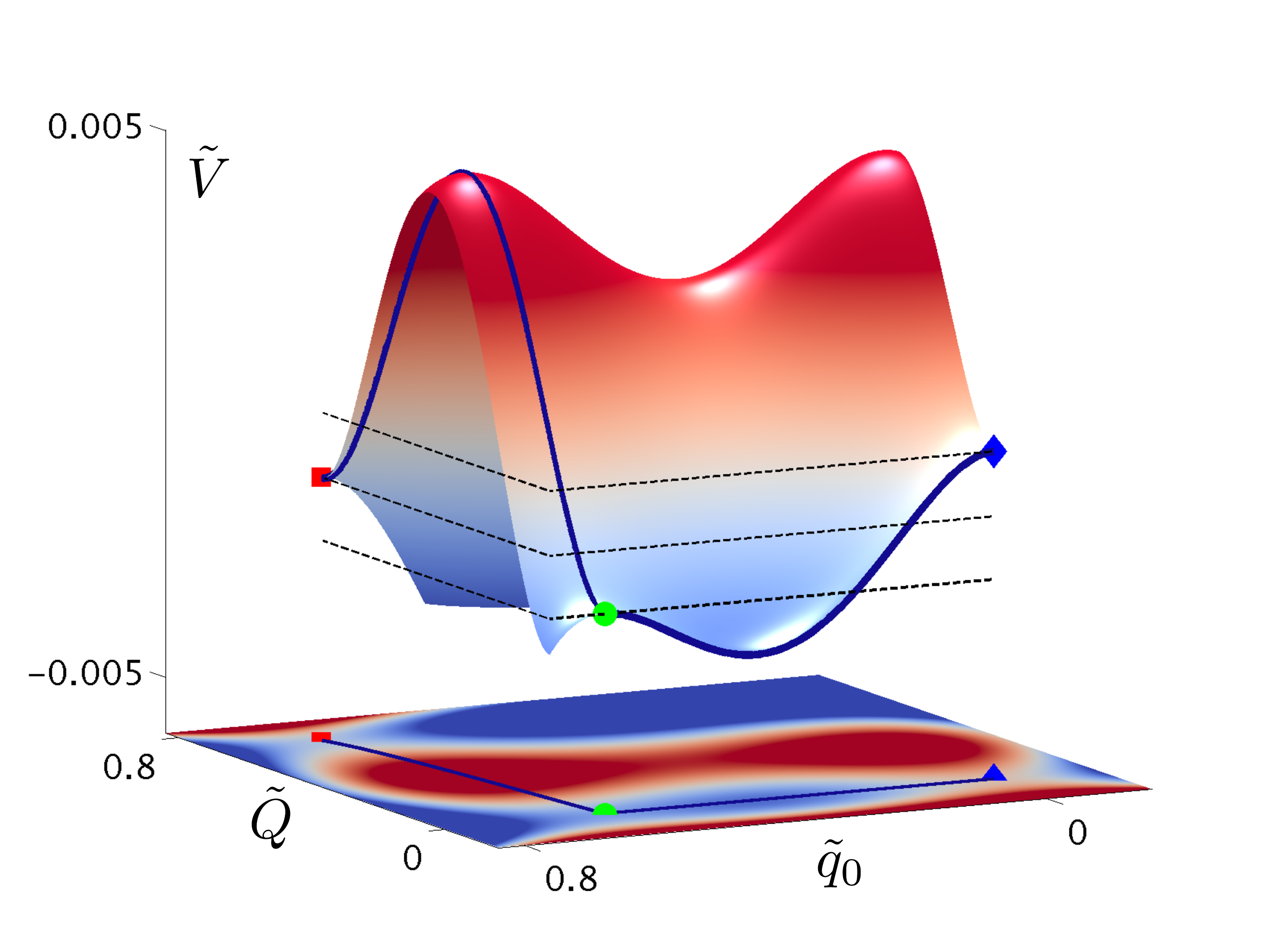}\label{instantonpath}}
\caption{
\textbf{(1+0)-step RSB instantons.}
We display (1+0)-step RSB instantons at $\tilde{t}\equiv t\, y/w^2=0.217$ and $\tilde{u}\equiv u/w=0.04$, well below the yellow dashed line of Fig.~\ref{stabilities} where the instanton develops a two-walled structure.
In (a) we display dimensionless fields $\{\tilde{\Q},\tilde{\qzero}\}\equiv\tfrac{y}{w}\{\Q,\qzero\}$ as a function of the dimensionless coordinate $\tilde{r}\equiv r\, w/\sqrt{y}$.
In (b) the solid line indicates the trajectory of the same configuration along the dimensionless potential $\tilde{V}=V(\tilde{\Q},\tilde{\qzero}) \, y^3/w^4$.
The red square, green sphere, and blue diamond indicate the metastable, transient, and trivial states.
The initial excursion proceeds as usual, connecting the metastable extremum to the lower-$\tilde{V}$ transient extremum by passing over the intervening potential barrier.
The subsequent excursion, however, is \emph{inverted} due to the ``wrong" sign kinetic term, connecting the transient extremum to the higher-$\tilde{V}$ trivial extremum by passing over an inverted potential barrier.}
\label{instantons}
\end{figure*}

\begin{figure}[t]
\centering
{\includegraphics[width=8cm]{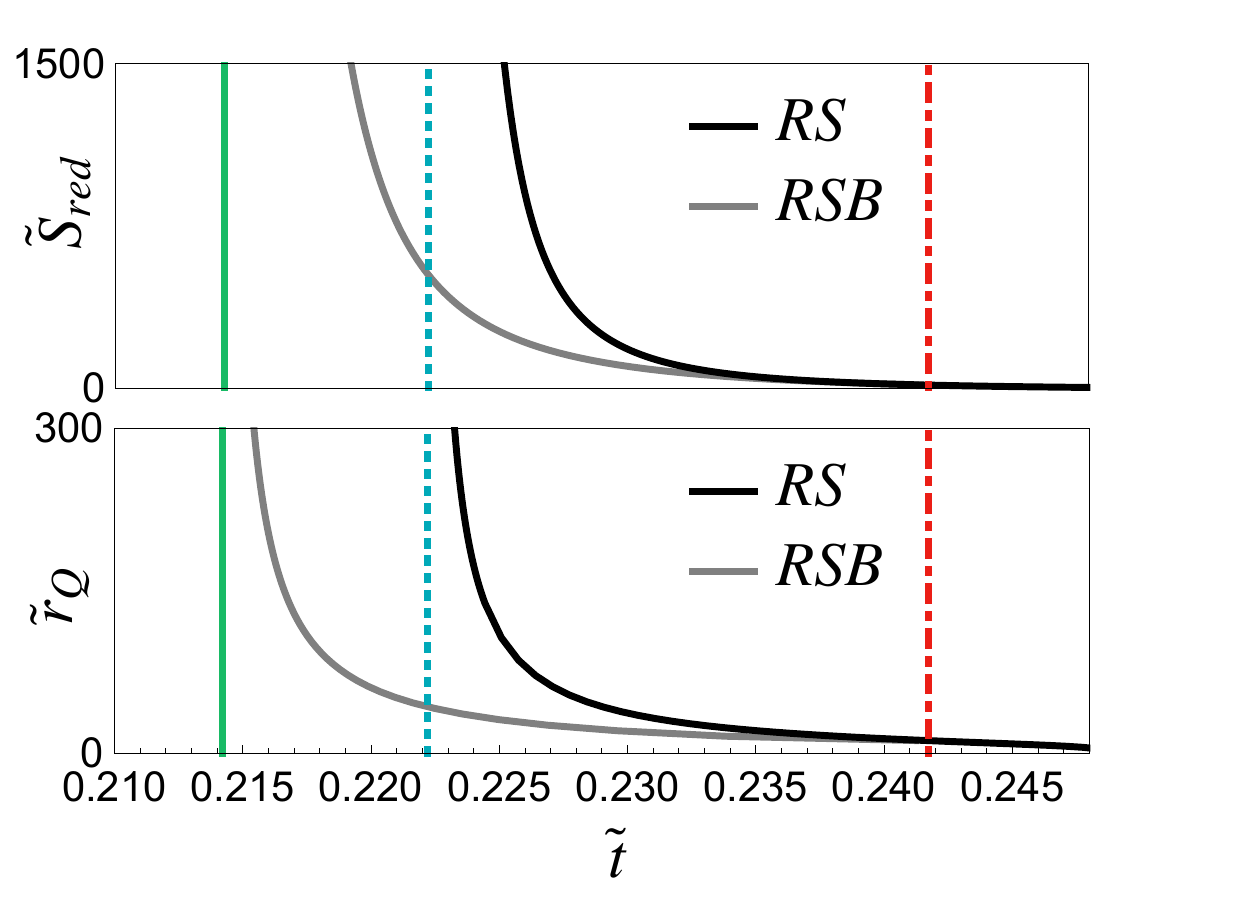}}
\caption{
\textbf{Instanton sizes and actions.}
At fixed $\tilde{u}\equiv u/w=0.04$, we show the dimensionless wall sizes $\tilde{r}_{\Q}\equiv r_{\Q}\, w/\sqrt{y}$ and, reduced actions $\tilde{S}_{\rm red}\equiv S_{\rm red} y^{3/2}/w$ for the RS (black) and RSB (gray) instantons as a function of $\tilde{t}\equiv t\, y/w^2$.
The red dash-dotted line indicates where the RS instanton develops an instability towards (1+0)-step RSB.
The blue dotted and green solid lines mark the critical values of $\tilde{t}$ for the RS and (1+0)-step RSB instantons.
To obtain a finite action, we shift the potential by a constant such that the homogeneous metastable solution, $\Q\le({\bf r}\ri)= Q_{\rm meta}$, has zero action.}
\label{divergence}
\end{figure}

Above the curve in Fig.~\ref{stabilities}  upon which the RS instanton becomes unstable, the only solution to the modified instanton Eqs.~(\ref{N0RSB_EOM_1}) and (\ref{N0RSB_EOM_2}) is the original RS instanton with $\qzero\le({\bf r}\ri)=Q\le({\bf r}\ri)$.
Along this curve, another branch of solutions appears with $\qzero\le({\bf r}\ri)\neq Q\le({\bf r}\ri)$.

Various features of these instantons are captured by the reduced action
\bea\label{N0RSB}
&&S_{\rm red}\le[\Q\le({\bf r}\ri),\qzero\le({\bf r}\ri)\ri]\equiv\lim_{\nr\rightarrow1}\frac{2 S[q_{ab}\le({\bf r}\ri)]}{\nr(\nr-1)}\Big|_{\rm (1+0)  \ RSB}\nonumber\\
&=&\int d^3{\bf r}\ \le[\le(\nabla \Q\ri)^2-\frac{1}{2}\le(\nabla \qzero\ri)^2+V\le(\Q,\qzero\ri)\ri]
\eea
where
\bea\label{crazypot}
V\le(\Q,\qzero\ri)&=&2\le[\frac{t}{2}\Q^2-\frac{w}{3}\Q^3+\frac{y}{4}\Q^4\ri]\nonumber\\
&&-\le[\frac{t}{2}\qzero^2-\frac{w}{3}\qzero^3+\frac{y}{4}\qzero^4\ri]\nonumber\\
&&-\frac{u}{3}\le[2\Q^3+\qzero^3-3\Q^2 \qzero\ri]\, .
\eea
Note that the kinetic term for $q_0$ has the wrong sign, as do the pure-$q_0$ terms in the potential.
This sign follows from the number of $q_0$ elements in Eq.~(\ref{N0RSBpic}) being $(N_{\rm r}-1)(N_{\rm r}-2)/2$, which formally goes negative in the $N_{\rm r} \to 1$ limit of Eq.~(\ref{N0RSB}).
This sign-flipping is familiar from the study of mean-field spin-glasses and does not pose any serious trouble; here, for on-shell configurations satisfying the saddle-point equations, large field excursions are suppressed in both $Q$ and $q_0$.
Nonetheless, this sign plays an important role in what follows.

The (1+0)-step RSB instanton asymptotes to the metastable state $(\Q,\qzero)=(Q_{\rm meta},Q_{\rm meta})$ at large $r$ and approaches the trivial state $(\Q,\qzero)=(0,0)$ in the interior near $r=0$.
In certain parameter regimes, the instanton develops a two-walled structure in which the fields linger near a metastable ``transient state,''
\be\label{trans}
(\Q,\qzero)=\le(0,Q_{\rm tran}\equiv\frac{(w-u)+\sqrt{(w-u)^2-4yt}}{2y}\ri)\, ,
\ee
over an intermediate range of $r$ (Fig.~\ref{instantonsb}).
This transient state exists for $4yt<(u-w)^2$, in other words below the yellow dashed line in Fig.~\ref{stabilities}.
Well below this curve, the (1+0)-step RSB instanton develops the two-walled structure.

Importantly, the size of the outer wall diverges \emph{not} at the critical value for the RS instanton $t^{\rm RS}_{\rm c}=\frac{2w^2}{9y}$, but at a different critical line (see Figs.~\ref{stabilities} and~\ref{divergence}).
This divergence can be understood by generalizing the standard thin-wall argument to the reduced potential (\ref{crazypot}).
In essence, as we tune parameters such that the metastable potential energy $V(Q_{\rm meta},Q_{\rm meta})$ approaches the transient potential energy $V(0,Q_{\rm tran})$, the size of the outer wall diverges.
It is worthwhile to mention that there exists a path of finite-action from the metastable state to the trivial state even when the metastable potential energy is smaller than the trivial potential energy $V(0,0)$.
This is a legacy of the curious signs in the reduced effective action. We depict a trajectory in field space for a typical two-walled instanton in Fig.~\ref{instantonpath}.

As a further check of our stability analysis, we have confirmed that the (1+0)-step RSB instanton, when it exists, has a smaller wall size and reduced action than the RS instanton (see Fig.~\ref{divergence}).
This implies that the (1+0)-step RSB instanton dominates over the RS instanton in determining the PTS correlations.

The method outlined above extends naturally to higher-step RSB instantons.
Carrying out the stability analysis around the (1+0)-step RSB instanton, we indeed find an instability towards a (1+1)-step RSB instanton.
Schematically,
\be\label{1+1RSB}
q_{ab}\le({\bf r}\ri)=\left[ {\begin{array}{cccccc}
   0 & \Q & \Q & \Q &\Q & ...\\
   \Q & 0 & \qone & \qzero & \qzero & ... \\
   \Q & \qone & 0 & \qzero & \qzero & ... \\
    \Q & \qzero & \qzero & 0 & \qone & ... \\
    \Q & \qzero & \qzero & \qone & 0  & ...\\
   ... & ... & ... & ... & ... & ... \\
  \end{array} } \right]\le({\bf r}\ri)\, .
\ee
This means that further RSB proceeds by breaking the residual replica symmetry of $q_{\tilde{a}\tilde{b}}\le({\bf r}\ri)=(1-\delta_{\tilde{a}\tilde{b}})\qzero\le({\bf r}\ri)$ in the mean-field spin-glass way~\cite{Parisi}, in agreement with the picture that a new configuration inside the cavity is immersed in a disorder created by the original.
Extending this analogy to spin glasses, we expect (1+1)-step RSB to entail hierarchical clustering of metastable states in the free-energy landscape for particles inside the cavity, with concomitant changes in dynamics~\cite{Stillinger}.
A complete analysis will be reported elsewhere, but it is worth mentioning here that spherical symmetry is also broken at this stage.
This is in agreement with the recent study carried out in a wall point-to-set geometry~\cite{BCnew}, which suggests that cooperatively rearranging regions have rough interfaces with a growing ``wandering length."
All in all, our work suggests that the PTS correlations contain much richer information than just a length scale, and generic RSB patterns and their implications for glassy dynamics should be fully explored.\\

{\bf Acknowledgments}
We thank Francesco~Zamponi for enlightening discussions.
A.A., T.A.~and S.Y.~are supported in part by the U.S.~Department of Energy (DOE) under cooperative research agreement No.~DE-FG02-05ER41360 and J.L.~is supported in part by Samsung Scholarship.
S.Y.~is also supported in part by the Alfred P.~Sloan Foundation and the European Research Council under the European Union's Seventh Framework Programme (FP7/2007-2013)/ERC Grant Agreement No.~306845.
\\

\appendix*

\section{}

\subsection{Pseudospectral method}

We detail here the numerical methods used to solve our BVPs, focusing first on the RS instanton and subsequently explaining how the specific details change when considering instantons in the (1+0)-step RSB sector.
All the computations were coded in MATLAB and carried out on desktop computers.\\

{\it RS instanton} --
We must solve the nonlinear saddle-point equation
\be\label{trivialEOMA}
-\nabla^2\Q+t\Q-w\Q^2+y\Q^3 = 0
\ee
with the Neumann boundary condition $\tfrac{dQ}{dr}\big|_{r=0}=0$ and the Dirichlet boundary condition $Q(\infty)=Q_{\rm meta}$.
To solve this BVP efficiently, we represent $\Q(r)$ pseudospectrally in a basis of $N$ Chebyshev polynomials, $T_k(x(r))$, keeping track of the field variables at the Chebyshev extrema collocation grid.  Here, $x(r)=b_0\tanh\{\alpha_Q(r-r_\Q)\}+b_1$, with $b_0$ and $b_1$ chosen such that the domain $r\in [0,\infty]$ maps onto the compact interval $x\in [-1,1]$.
The parameter $r_{\Q}$ is a proxy for the position of the wall, implicitly defined by $\Q(r_\Q)\equiv\Q_{\rm meta}/2$. 
We choose the remaining parameter $\alpha_{\Q}$ such that the Chebyshev collocation points are well concentrated around the instanton wall where the derivatives of $\Q(r)$ are expected to be large: the fixed choice {$\alpha_{\Q}=0.1 w/\sqrt{y}$} was sufficient for our needs. 
Once the coordinate parameters are chosen, we solve the nonlinear equations for field values at the collocation points using Newton's method~\cite{SpectralMatlab}.
This requires an initial guess that, if well chosen, will converge to the desired solution.

The coordinate parameter $r_{\Q}$  is adjusted as computations progress.
Namely, given a guess $\Q_{\rm g}$, we read off its associated wall location $r_{\Q_{\rm g}}$.
We then use it to fix the coordinate and refine the solution using Newton's method.
The result {$\Q_{\rm s}$} generically has a new wall location $r_{\Q_{\rm s}}$, and we use this to readapt the coordinates.
This is iterated until the coordinate parameter converges.

The Newton iteration is extremely sensitive to the proximity of the guess to the RS instanton, with bad guesses converging to the homogeneous solutions.
To surmount this problem, we use {a ``mountain pass" algorithm, which forces the field to traverse the saddle of the potential as follows.}
We first make an initial guess of the position of the instanton wall, which we call {$r_{\rm g}$}.
We then divide the full domain into two, $[0,r_{\rm g}]$ and $[r_{\rm g},\infty]$, and we solve a different BVP in each region.
Specifically, in the region $[0,r_{\rm g}]$, we solve the saddle-point equation with mixed Neumann-Dirichlet boundary conditions $(d\Q/dr)\big|_{r=0}=0$ and $\Q(r_{\rm g})=Q_{\rm meta}/2$.
In the region $[r_{\rm g},\infty]$, we impose Dirichlet boundary conditions $\Q(r_{\rm g})=Q_{\rm meta}/2$ and $\Q(\infty)=Q_{\rm meta}$. 
For a generic choice of the floating parameter $r_{\rm g}$, the patched solution has a kink at $r=r_{\rm g}$.
We then vary $r_{\rm g}$ until the left- and right-sided first derivatives match. 
The patched solution then provides a guess that generally converges rapidly to the RS instanton on the whole domain $r\in [0,\infty]$.  The value of $r_{\rm g}$ provides an initial guess of $r_{\Q}$.
\\

{\it (1+0)-step RSB instantons} -- In this sector we must solve the equations,
\be\label{N0RSB_EOM_1A}
-\nabla^2\Q+t\Q-w\Q^2+y\Q^3 = -u\le[\,\Q\qzero-\Q^2\ri]
\ee
and
\be\label{N0RSB_EOM_2A}
-\nabla^2\qzero+t\qzero-w\qzero^2+y\qzero^3 = -u\le[\,\qzero^2-\Q^2\ri]\, ,
\ee
for two fields $\Q(r)$ and $\qzero(r)$, whose profiles must not coincide. The emerging two-walled structure for the field $\qzero(r)$ motivates the use of the following compact coordinate:
\be
x =b_0\le[\tanh\le\{\alpha_{\Q} (r -r_{\Q})\ri\}+\tanh\le\{\alpha_{0} (r -r_{0})\ri\}\ri]+b_1
\ee
{for $\qzero$}. Here $r_{\Q}$ and $r_{0}$ are proxies for the locations of the walls of the instantons, defined implicitly by $\Q(r=r_\Q)=\qzero(r=r_0)=Q_{\rm meta}/2$ and weakly adapted.
Again $\alpha_{\Q}$ and $\alpha_{0}$ are set to $0.1 w/\sqrt{y}$ . 

Applying the mountain pass algorithm in this sector, we start with a guess for $r_\Q$, again called $r_{\rm g}$, \emph{and} the value of $\qzero(r_{\rm g})\equiv\qzero^{\rm g}$.
We then impose $\Q(r_{\rm g})=Q_{\rm meta}/2$ and $\qzero(r_{\rm g})=\qzero^{\rm g}$ at the patching point, varying  the floating parameters $r_{\rm g}$ and $\qzero^{\rm g}$ until the first derivatives match at either side of the patching point.
To  avoid the RS solution for which $\Q(r)=\qzero(r)$, we conduct this search with the restriction $\qzero^{\rm g}\neq Q_{\rm meta}/2$.

This algorithm is reliable in an open neighborhood of the instability line.
There we may gradually increase $\qzero^{\rm g}$ away from the RS value $Q_{\rm meta}/2$ until a smooth RSB instanton is reached.
Once such a solution is found and polished on the whole domain, we can efficiently explore the parameter space by adiabatically shifting the solution: given a solution at some point in parameter space, we can use it as a guess for  adjacent points.

Our results are robust against changes in the number of collocation points, as long as it is large enough.
Specific data used in this paper are generated with $41$ collocation points for the field $\Q$ and $81$ collocation points for the field $\qzero$.\\

\subsection{Eigenvalue spectrum}
The differential operator of interest, defined on the left-hand side of equation
\be\label{secondA}
\le[-\nabla^2+t-\le(2w-u\ri)\Q^\star+3y\Q^{\star 2}\ri]\delta\phi_{\rm II}^{(n)}=\lambda_{\rm II}^{(n)}\delta\phi_{\rm II}^{(n)}\, ,
\ee
is spherically symmetric.
Thus we can first diagonalize the operator in the spherical directions by expanding in spherical harmonics.
Then, for each harmonic, we can numerically obtain the eigenspectrum by viewing the operator as a matrix acting on the projected space of regular functions with definite angular momentum.\\


\begin{thebibliography}{11}
\bibitem{footnoteparameter}
For simplicity, we restrict our analysis to the parameter region $0\leq u\leq0.05 w$, with $w,y>0$.
\bibitem{footnotenaming}
We refer to this as (1+0)-step RSB since this first step is qualitatively differente from what one is accustomed to from spin-glasses.  Higher-order breaking, however, appears to be conventional, so we shall refer to such higher-order RSB  as  (1+$k$)-step RSB.
\end{thebibliography}
\end{document}